\documentclass[
    11pt,
    letterpaper,
    reprint,
    nofootinbib,
    notitlepage,
    superscriptaddress,
    aps,prl
]{revtex4-2}

\usepackage{amsmath,amssymb,amsfonts} 
\usepackage{empheq}
\usepackage{physics}

\usepackage[svgnames,dvipsnames]{xcolor}
\usepackage{graphicx}
\definecolor{NewBlue}{rgb}{0.1, 0.1, 0.7}
\definecolor{NewRed}{rgb}{0.7, 0.1, 0.1}
\usepackage[colorlinks,
    linkcolor=Maroon,
    citecolor=NewBlue,
    urlcolor=NewRed]{hyperref}

\usepackage{cleveref}

\usepackage{dcolumn}
\usepackage{bm}
\usepackage{dsfont}
\usepackage[normalem]{ulem}

\renewcommand{\t}[1]{\mathrm{#1}}
\newcommand{\avg}{\expval}

\begin{document}

\title{Causal State Estimation and the Heisenberg Uncertainty Principle}

\author{Junxin Chen}
\email{junxinc@mit.edu}
\author{Benjamin B. Lane}
    \affiliation{LIGO, Massachusetts Institute of Technology, Cambridge, MA 02139, USA}
\author{Su Direkci}
    \affiliation{
    The Division of Physics, Mathematics and Astronomy, California Institute of Technology, CA 91125, USA
    }
\author{Dhruva Ganapathy}
\author{Xinghui Yin}
\author{Nergis Mavalvala}
    \affiliation{LIGO, Massachusetts Institute of Technology, Cambridge, MA 02139, USA}
\author{Yanbei Chen}
    \affiliation{
    The Division of Physics, Mathematics and Astronomy, California Institute of Technology, CA 91125, USA
    }
\author{Vivishek Sudhir}
\email{vivishek@mit.edu}
\affiliation{LIGO, Massachusetts Institute of Technology, Cambridge, MA 02139, USA}
\affiliation{
 Department of Mechanical Engineering, 
 Massachusetts Institute of Technology, Cambridge, MA 02139, USA
}

\date{\today}

\begin{abstract}
The observables of a noisy quantum system can be estimated by appropriately filtering the 
records of their continuous measurement. Such filtering is relevant for state estimation, and if the 
filter is causal, also relevant for measurement-based feedback control. 
It is therefore imperative that a pair of conjugate observables
estimated causally satisfy the Heisenberg uncertainty principle. 
In this letter, we prove this fact --- without assuming Markovian dynamics 
or Gaussian noises, in the presence or absence of feedback control of the system, and where in the 
feedback loop (inside or outside) the measurement record is accessed. Indeed, causal estimators
using the in-loop measurement record can be as precise as those using the out-of-loop record.
These results clarify the role of causal estimators to non-Markovian quantum systems, 
restores the equanimity of in-loop and out-of-loop measurements in their estimation and control, and 
simplifies future experiments on measurement-based quantum feedback control.
\end{abstract}

\maketitle

\emph{Introduction. }
Estimating the state of a system from a noisy measurement record is ubiquitous in engineering and
science. For a classical linear system driven by stationary Gaussian noise, the optimal estimator ---
in the sense of minimizing the mean-square error --- is given by the Wiener filter
\cite{Wiener42,Kailath1981}. For prediction problems where only the past measurement record is
available, the Wiener filter is causal and is given in terms of the spectrum of the measurement
record and a model of its cross-spectrum with the state. 
Additionally, if the dynamics are Markovian, a state-space model of the system 
and its observation is available and the Kalman filter uses that description to produce an equivalent, 
more tractable estimate of the state \cite{Kalman60}.
State estimation is also a crucial element in optimal control: the separation principle
\cite{Won68,Lindq73} asserts that the feedback controller that minimizes a quadratic cost
function of a linear system driven by Gaussian noise can be split into a causal state estimator
and a linear regulator, with any error in the controlled state set by that of the estimator.

These ideas have been fully transposed to quantum systems: the theory of quantum state estimation
\cite{Bela89,BarchBela91,Bela92,Wise96,Doherty1999,BoutMaas04}, feedback control
\cite{wiseman1994quantum,WiseDoh05,GenSer13,WiseMil,Jacobs}, and the separation principle
\cite{BoutHand08} have been developed. A central distinction between classical and quantum
systems is that the latter has to obey Heisenberg's uncertainty principle: 
a pair of conjugate observables cannot be determined with arbitrary precision simultaneously, even in principle. 
Consequently, it is believed that causal estimation of conjugate observables cannot be simultaneously
more precise than that allowed for the observables themselves. 
Violation of this expectation would result in an unphysical conditional state 
(i.e. a state estimated through the measurement record) \cite{Doherty1999}, and,
by the separation principle, the possibility of 
transmuting the unphysical conditional state to an unphysical steady state by feedback control. 

The purpose of this paper is to prove, in general and explicitly, that the 
product of the variance of the errors in the causal estimate of observables 
is lower bounded by the minimum variance product of corresponding physical observables allowed by 
the Heisenberg uncertainty principle.
In another words, the uncertainties of causal estimation errors respect the uncertainty principle of the corresponding physical observables, and thus the causal conditional state has to be physical.
Prior work guarantees this in Markovian and Gaussian settings through 
\cite{Bela89,BarchBela91,Bela92,Doherty1999,Tsang2009Quantum}.
However, recent experiments in quantum state estimation and feedback control in systems ranging 
from atoms \cite{HoodKimb00,KubRemp09,SayHar11,ZhoHar12}, 
solid state qubits \cite{VijSid12,HatDev13,murch2013observing,weber2014mapping}, 
to mechanical oscillators
\cite{wilson2015measurement,HacSid16,SudKip17,rossi2018measurement,Rossi2019,minev2019catch,
whittle2021approaching,tebbenjohanns2021quantum,magrini2021real}
call for a simple and general guarantee that causal state estimation and control of quantum
systems will not violate the basic tenet of the uncertainty principle, including non-Markovian cases. 
In particular, non-Markovian $1/f$ noise limits coherence of  
solid-state mechanical oscillators \cite{GonSaul95,Kaj99,groeblacher2015observation,NebMav12,FedKip18}
and superconducting qubits \cite{WellClar87,YosTsai06,BialMart07,KumDerm16}, both leading candidates for
measurement-based feedback control. In the following we address the non-Markovian setting generally.


\emph{Statement of problem and main result. }
Consider a quantum system, and any pair of observables $\hat{A}(t),\hat{B}(t)$, in the Heisenberg 
(or interaction) picture, 
which will be the object of estimation via continuous linear measurement.
The uncertainities in them are lower-bounded by 
the uncertainty principle
\begin{equation}\label{eq:RS}
\begin{split}
    \sigma_{A(t)}^2 \sigma_{B(t)}^2 
    \geq &\abs{\frac{1}{2}\avg{\comm{\hat{A}(t)}
        {\hat{B}(t)}}}^2 \\
          &+ \abs{\frac{1}{2} \avg{\acomm{\hat{A}(t)}{\hat{B}(t)}}}^2 \\
    \geq& \abs{\frac{1}{2}\avg{\comm{\hat{A}(t)}{\hat{B}(t)]}}}^2,
\end{split}
\end{equation}
where $\sigma_O^2 \equiv \avg*{\hat{O}^2}$ is the variance of the operator $\hat{O}$,
$\comm*{\cdot}{\cdot}$ is the commutator, and $\acomm*{\cdot}{\cdot}$ the anti-commutator.
Here we assume, without loss of generality, that observables are zero-mean.
In passing to the last line, we have also dropped an overall positive term on the right-hand side,
which makes the resulting bound weaker, but has the advantage that for bosonic 
(and linearized fermionic \cite{Klein91,Au94,yuan2022quantum}) systems the lower bound can be state-independent.

The premise of the uncertainty principle is the positivity of the quantum state $\hat{\rho}$:
clearly, $\t{Tr}[\hat{M}(t)^\dagger \hat{M}(t)\hat{\rho}] \geq 0$ for any operator $\hat{M}(t)$;
in particular, also for $\hat{M} = \hat{A} + \lambda \hat{B}$ for any constant $\lambda$. 
Further, the inequality must still hold for the minimum value of its left-hand side as a function of $\lambda$; 
this gives \cref{eq:RS}. 
So as long as the quantum state is guaranteed to be positive, the uncertainty 
relation holds. The problem is that in non-Markovian settings --- i.e. where the system may be driven 
by non-Markovian noises, its measurements may be contaminated by non-Markovian noises, and/or feedback 
maybe non-Markovian --- positivity of the conditional state is difficult to extract from models of its evolution.

We therefore analyse the
problem in the interaction picture for the observables of interest, assuming only linearity of measurement
and feedback. In particular, we will show that for causal linear measurement, the estimation errors
$\Delta \hat{A}, \Delta \hat{B}$ (to be defined below) of the observables $\hat{A},\hat{B}$
satisfy 
\begin{equation}
	\sigma_{\Delta A(t)}^2 \sigma_{\Delta B(t)}^2 
        \geq \abs{\frac{1}{2} \avg{\comm{\hat{A}(t)}{\hat{B}(t)}}}^2;
\end{equation}
i.e. causal estimates of conjugate observables are no more precise than the observables themselves.

\begin{figure}[t!]
    \centering\includegraphics[width=0.85\linewidth]{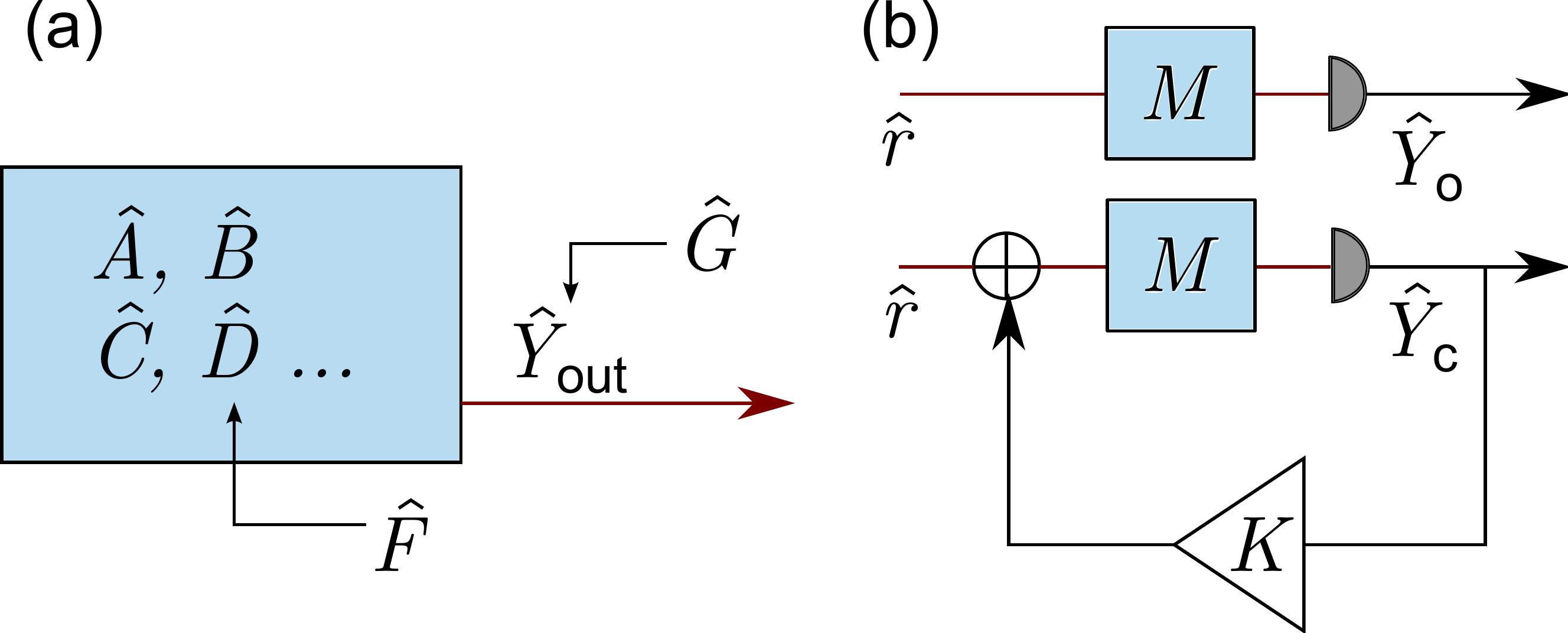}
    \caption{(a) Schematic diagram for a general open-loop measurement. $\hat{F}$ is a general force coupling to the system through system parameter $\hat{D}$. The effect of this interaction is encoded in the outgoing field $\hat{Y}_\mathrm{out}$ together with other system information. A general force $\hat{G}$ couples to the system through $\hat{Y}_\mathrm{out}$ does not influence the system due to causality. (b) Schematic diagram for an open-loop system (upper panel) and a measurement-based feedback controlled system (lower panel). $M$ and $K$ stand for the system and controller transfer functions respectively. $\hat{r}$, $\hat{Y}_\mathrm{o}$ and $\hat{Y}_\mathrm{c}$ are the set point, the open-loop measurement record and the in-loop measurement record respectively.}
    \label{fig:meas}
\end{figure}

\emph{Proof for case without feedback. }
We now consider a setup for quantum state estimation where the system is monitored continuously, 
the result of which is described by the measurement record $\hat{Y}$ (\Cref{fig:meas}(a)).  
The continuous monitoring condition is that \cite{BragKhal}
\begin{equation}\label{eq:yCont}
    [\hat{Y}(t),\hat{Y}(t')] = 0 \quad \t{for\; all}\quad t,t'.
\end{equation}
The observables are estimated by filters $W_{A,B}$ acting linearly on the record: 
\begin{equation}\label{eq:xpEst}
    \hat{A}_e(t) = (W_A*\hat{Y})(t), \quad \hat{B}_e(t) = (W_B*\hat{Y})(t),
\end{equation}
where $*$ stands for convolution. When $W_{A,B}$ are causal, i.e.
\begin{equation}\label{Hcausal}
	W_{A,B}(t\leq 0) = 0,
\end{equation}
they \emph{only} act on the past record. 

\Cref{eq:yCont} implies that the estimators commute with each other, 
and so they do not need to obey the uncertainty principle. 
That is, the uncertainty in the conditional state (i.e. the state estimated based on the measurement record) is not contained 
in the estimators, but in the estimation errors, described by the operators \cite{Doherty1999}
\begin{align}
    \Delta \hat{A}(t) &= \hat{A}(t) - \hat{A}_e(t) = \hat{A}(t)-(W_A*\hat{Y})(t) \label{eq:dx}\\
    \Delta \hat{B}(t) &= \hat{B}(t) - \hat{B}_e(t) = \hat{B}(t)-(W_B*\hat{Y})(t)\label{eq:dp},
\end{align}
which characterize how close the estimators follow the physical operators. 
To see this, note that in the Heisenberg picture, the positivity of the \emph{initial} state suffices to 
guarantee that
\begin{equation}\label{DADBuncert}
    \begin{split}
    \sigma_{\Delta A(t)}^2 \sigma_{\Delta B(t)}^2 
    \geq 
    \abs{\frac{1}{2}\avg{\comm{\Delta \hat{A}(t)}{\Delta \hat{B}(t)}}}^2.
    \end{split}
\end{equation}
That is, the estimation errors cannot be determined simultaneously with arbitrary precision.

We will now show that causality of the estimation filters $W_{A,B}$ [\cref{Hcausal}] and of the 
measurement implies
that $\avg*{\comm*{\Delta \hat{A}(t)}{\Delta \hat{B}(t)}} = \avg*{\comm*{\hat{A}(t)}{\hat{B}(t)}}$, 
so that the lower bound in \cref{DADBuncert} becomes the lower bound in the uncertainty principle for
the observables $\hat{A},\hat{B}$.

Using \cref{eq:dx,eq:dp} together with \cref{eq:yCont}, we have
\begin{align}\label{eq:commerr}
    \avg{[\Delta \hat{A}(t),\Delta\hat{B}(t)]}
    =& \avg{[\hat{A}(t),\hat{B}(t)]} - \avg{[\hat{A}(t), \hat{B}_e(t)]} \nonumber\\&- \avg{[\hat{A}_e(t),\hat{B}(t)]}. 
\end{align}
The key in determining the remaining average of equal-time commutators is an understanding of the 
\emph{unequal-time} commutators between the system observable and the measurement record.
In the context of state estimation and measurement-based quantum control, weak measurement applies to most of the current experiments \cite{HoodKimb00,KubRemp09,VijSid12,HatDev13,murch2013observing,weber2014mapping,wilson2015measurement,HacSid16,SudKip17,rossi2018measurement,Rossi2019,minev2019catch,tebbenjohanns2021quantum,magrini2021real}, thus we treat the effect of measurement on the system as a perturbation, and apply 
linear response theory \cite{kubo1957statistical,BragKhal,buonanno2002signal,Clerk04}. 
The key result of the theory is Kubo's formula, which gives the linear response of the average of a system observable $\avg*{\hat{C}(t)}$ to an external perturbation $\hat{F}(t)$, whose effect 
is described by the Hamiltonian $\hat{H}(t) = \hat{H}_0 - 
\hat{D}\hat{F}(t)$, where $\hat{H}_0$ is the free Hamiltonian and $\hat{D}$ is a system observable.
Kubo's formula states
\begin{equation}\label{eq:Kubo0}
    \avg{\hat{C}(t)} = \avg{\hat{D}^{(0)}(t)} + \frac{i}{\hbar} \int_{-\infty}^t \chi_{CD}(t,t') 
    \avg{\hat{F}(t')} \dd{t'}
\end{equation}
where $\hat{C}^{(0)}(t)$ is the observable evolving under $\hat{H}_0$ and
\begin{equation}\label{eq:Kubo1}
    \chi_{CD}(t,t') = \avg{\comm{\hat{C}^{(0)}(t)}{\hat{D}^{(0)}(t')}}
\end{equation}
is the linear response susceptibility, evaluated on the state of the system evolving 
under $\hat{H}_0$ (i.e. Kubo's formula holds in the interaction picture).

Consider now the schematic diagram of a general measurement as shown in \cref{fig:meas}(a) where
$\hat{Y}_\mathrm{out}=\hat{Y}$ is the \textit{outgoing} measurement record and $\hat{G}$ is the
generalized force coupling to the system through $\hat{Y}_\mathrm{out}$ by the interaction
Hamiltonian $-\hat{Y}_\mathrm{out}\hat{G}$. Focusing on $\hat{A}$ and using \cref{eq:Kubo0}, we have
\begin{equation}
    \avg{\hat{A}(t)}=\avg{\hat{A}^{(0)}(t)}+\frac{i}{\hbar}\int_{-\infty}^t 
     \chi_{AY}(t,t')\avg{\hat{G}(t')}\dd{t'},
\end{equation}
with $\chi_{AY}(t,t')=\avg*{[\hat{A}^{(0)}(t),\hat{Y}_\t{out}^{(0)}(t')]}$.
Now we bring in the crucial ingredient of causality of the measurement: 
for an open-loop measurement, i.e. where the measurement record is not used for feedback, 
the measurement record $\hat{Y}$ cannot influence the system dynamics at a later time, 
implying that:
\begin{equation}
    \chi_{AY}(t,t')=\avg{[\hat{A}^{(0)}(t),\hat{Y}^{(0)}(t')]}=0 \qfor t\geq t'.
\end{equation}
Since the susceptibility is zero for $t \geq t'$, the force $\hat{G}$ does not influence the system for
such times. Thus
\begin{equation}\label{eq:commxy}
    \chi_{AY}(t,t')=\avg{[\hat{A}(t),\hat{Y}(t')]}=0 \qfor t\geq t',
\end{equation}
where we have dropped the superscripts ``(0)".

Let us now consider $\avg*{[\hat{A}(t),\hat{B}_e(t)]}$. Without loss of generality, we evaluate it in the 
interacting picture, where the arbitrary state sandwiching the commutator evolves under $H_0$. 
Using \cref{eq:xpEst},
\begin{equation}
    \avg{[\hat{A}(t), \hat{B}_e(t)]} = \int_{-\infty}^{+\infty} W_B (\tau) \avg{[\hat{A}(t),\hat{Y}(t-\tau)]} 
    \dd{\tau}.
\end{equation}
By causality of the estimation filter, $W_B(\tau) = 0$ for $\tau < 0$. For $\tau \geq 0$, \cref{eq:commxy} implies 
that $\avg*{[\hat{A}(t), \hat{Y}(t-\tau)]} = \chi_{AY}(t,t-\tau) = 0$. 
Thus causality --- of the measurement interaction and 
of the estimation filter --- implies that $\avg*{[\hat{A}(t), \hat{B}_e(t)]} = 0$. Similarly, it can be 
shown that $\avg*{[\hat{A}_e(t), \hat{B}(t)]} = 0$. 

In summary, for causal estimation using open-loop measurements,
\begin{equation}\label{eq:commOL}
    \avg{[\Delta \hat{A}(t), \Delta \hat{B}(t)]}=\avg{[\hat{A}(t), \hat{B}(t)]},
\end{equation}
i.e., the estimation errors respect the uncertainty principle of the corresponding physical observables.

\emph{Proof for case with feedback. }
We now consider the case where the system is feedback-controlled, and using the in-loop 
measurement record $\hat{Y}_\t{c}$ for state estimation (see the lower panel of \cref{fig:meas}(b)). In this case the
in-loop record does affect the system at later times after the measurement interaction.
However the open-loop record $\hat{Y}_\t{o}$ does not.
In order to use this fact, note that the in-loop and open-loop measurement records, 
are related to each other as
\begin{equation}\label{eq:ycyo}
    \hat{Y}_\t{c}(t) = (K_\t{c} * \hat{Y}_\t{o})(t),
\end{equation}
where $K_\t{c}(t)$ is the inverse Fourier transform of $(1- M[\omega] K[\omega])^{-1}$ 
($M$ and $K$ are the transfer functions of the system and the controller respectively, as in \cref{fig:meas}(b).)
Importantly, if $M,K$ are causal, then so is $K_\t{c}$. 

Imagine now estimation based on the in-loop measurement record $\hat{Y}_\t{c}$: 
$\hat{A}_\mathrm{ec}(t)=(W_A^\t{c}*\hat{Y}_\mathrm{c})(t)$ and $\hat{B}_\mathrm{ec}(t)=(W_B^\t{c}*\hat{Y}_\mathrm{c})(t)$, with $W_A^\t{c}$ and $W_B^\t{c}$ the estimation filters in the in-loop case.
Following the same line of reasoning as in the open-loop case, the question of whether the estimated
observables respect the Heisenberg uncertainty principle of the original observables boils down to whether 
$\avg*{[\hat{A}(t), \hat{Y}_c(t')]} = 0$ for $t \geq t'$. Clearly, 
\begin{align*}
    \avg{[\hat{A}(t), \hat{Y}_c(t')]} 
    = \int_{-\infty}^{\infty} K_c (t' -\tau) 
    \avg{[\hat{A}(t), \hat{Y}_\t{o} (\tau)]} \dd{\tau}.
\end{align*}
We consider the integrand in two complementary intervals, $t'<\tau$ and $t'\geq\tau$. In the former region, $K_\mathrm{c}(t'-\tau)=0$ since $K_\mathrm{c}$ is causal. 
In the latter region, since $t\geq t'\geq\tau$, we have that 
$\avg*{[\hat{A}(t), \hat{Y}_\t{o} (\tau)]}=0$ using causality of open-loop dynamics. 
In sum, the integral is zero in both intervals due to causality.
Thus $\avg*{[\hat{A}(t), \hat{Y}_c(t')]}$ vanishes as long as $t\geq t'$. 
A similar argument holds for $\avg*{[\hat{B}(t), \hat{Y}_c(t')]}$. Thus
$\avg*{[\hat{A}(t),\hat{Y}_c(t')]} = 0 = \avg*{[\hat{B}(t), \hat{Y}_c(t')]}$ for $t\geq t'$.
In summary, the in-loop estimation errors
$\Delta \hat{A}_c = \hat{A} - \hat{A}_\t{ec}, \Delta \hat{B}_c = \hat{B}-\hat{B}_\t{ec}$
satisfy
\begin{equation}
    \avg{[\Delta \hat{A}_\mathrm{c}(t), \Delta \hat{B}_\mathrm{c}(t)]}=\avg{[\hat{A}(t), \hat{B}(t)]},
\end{equation}
just as in the open-loop case. 
Thus the in-loop causal estimation errors respect the uncertainty principle of the corresponding physical observables just as in the open-loop case. 

\emph{Effect of feedback on estimation error. }
The in-loop measurement record $\hat{Y}_\t{c}$ is often deemed untrustworthy for state 
estimation. Historically, this view originated from the apparent violation of the Heisenberg uncertainty 
principle by the in-loop field in linear measurement-based feedback control of 
optical fields \cite{YamMach86}. This behavior, 
called ``noise squashing'' \cite{wiseman1999squashed}, arises because fields inside a feedback 
loop are not freely propagating and so need not satisfy the canonical commutation relations 
\cite{shapiro1987theory,TaubWise95}.

In fact, if a model of the feedback loop is available, then state estimation using the 
in-loop measurement record is as accurate as an estimate based on the out-of-loop record.
We can always write the system observable in the presence of feedback, 
$\hat{A}_\mathrm{c}$, as the sum of the open-loop one $\hat{A}_\mathrm{o}$ plus a feedback 
term:
\begin{equation}
    \hat{A}_\mathrm{c}(t)
    =\hat{A}_\mathrm{o}(t)+\hat{A}_\mathrm{fb}(t)
    =\hat{A}_\mathrm{o}(t)+(K*\hat{Y}_\mathrm{c})(t),
\end{equation}
where $K$ is a causal feedback control filter. 
Now consider two filters $W_A^i [\omega]$ $(i=\t{c,o})$, one estimating the observable $\hat{A}$ from 
the in-loop record and the other from the out-of-loop record. 
The respective errors are $\Delta \hat{A}_i(t) = \hat{A}_i(t) - (W_A^i * \hat{Y}_i)(t)$.
It is straightforward to show that if the in-loop filter is chosen to be
\begin{equation*}
    W_A^\mathrm{c}[\omega] = \frac{W_A^\t{o}[\omega]}{K_\mathrm{c}[\omega]}+K[\omega] 
    = (1-W_A^\t{o}[\omega] M[\omega])K[\omega] + W_A^\t{o}[\omega],
\end{equation*}
then $\Delta \hat{A}_\t{c} = \Delta \hat{A}_\t{o}$.
Note that if the feedback loop and $W_A^\t{o}$ are stable and causal, so is $W_A^\t{c}$.
Therefore by proper filter design, state estimation can be performed using the 
in-loop measurement record without loss of fidelity.

\emph{Example of a structurally damped mechanical oscillator.} 
A canonical example of non-Markovian behavior of a subject of contemporary interest to quantum state estimation is a
structurally damped mechanical oscillator 
\cite{GonSaul95,Kaj99,groeblacher2015observation,NebMav12,FedKip18,whittle2021approaching}. 
Such a system is described by the linear response of its displacement $\hat{x}[\Omega]=\chi[\Omega]\hat{F}[\Omega]$,
to the force $\hat{F}$, by the susceptibility \cite{Saul90}
\begin{equation}\label{chi}
    \chi[\Omega] = [m(-\Omega^2 + \Omega_0^2 +i \Omega_0^2 \phi[\Omega])]^{-1},
\end{equation}
where $m$ is the mass, $\Omega_0$ is the resonance frequency, and $\phi$ is the loss angle. 
The thermal displacement fluctuations of the oscillator, as given by the fluctuation-dissipation theorem \cite{CalWel51,Kubo66}, 
has the spectrum
\begin{equation}\label{Sxx}
\begin{split}
    S_{xx}^\t{th}[\Omega] &= 2\hbar(n_\t{th}[\Omega]+\tfrac{1}{2})\Im \chi[\Omega] \\
        &= \frac{2\hbar (n_\t{th}[\Omega]+\tfrac{1}{2})}{m[(\Omega^2 -\Omega_0^2)+(\Omega_0 \phi[\Omega])^2]},
\end{split}
\end{equation}
with $n_\mathrm{th}[\Omega]=1/(\exp[\hbar\Omega/k_\mathrm{B}T]-1)$ being the thermal occupancy of the bath, 
where $k_\mathrm{B}$ is the Boltzmann constant, and $T$ the bath temperature. 

We will now show that a causal Wiener filter applied to an interferometric measurement of a structurally
damped oscillator produces estimates of its position $\hat{x}(t)$ and momentum $\hat{p}(t)$, whose estimation
errors satisfy
\begin{equation}\label{eq:upxp}
    \begin{split}
    \sigma_{\Delta x(t)}^2 \sigma_{\Delta p(t)}^2 
    &\geq \abs{\frac{1}{2} \avg{\comm{\Delta\hat{x}(t)}{\Delta\hat{p}(t)}}}^2\\
    &= \abs{\frac{1}{2} \avg{\comm{\hat{x}(t)}{\hat{p}(t)}}}^2
    = \frac{\hbar^2}{4}.
    \end{split}
\end{equation}

The structural damping model is unphysical and inconsistent if the loss angle $\phi[\Omega]$ is 
frequency independent. Unphysical, because the 
$1/\Omega$ scaling of the oscillator's displacement spectrum would preclude a
finite variance for the displacement. Mathematically inconsistent, because the susceptibility
between hermitian operators must have the symmetry $\chi[\Omega]^* = \chi[-\Omega]$, or equivalently,
$\Re\chi^{-1}[\Omega] =\Re \chi^{-1}[-\Omega]$ and $\Im \chi^{-1}[\Omega] = - \Im \chi^{-1}[-\Omega]$. 
For a structurally damped oscillator, $\Im \chi^{-1}[\Omega]=m\Omega_0^2 \phi[\Omega]$, is not
anti-symmetric in frequency if the loss angle is finite and frequency-independent. 
In order to satisfy the anti-symmetry, to leading order,
$\Im \chi^{-1}[\Omega \rightarrow 0] \propto \Omega$, which precisely cancels the $1/\Omega$ pathology in
the spectrum. 
The simplest example of such a loss angle is velocity-propotional damping, which however
is inconsistent with observations on high-quality elastic oscillators. 
The Zener model \cite{Zen40}
\begin{equation}
    \phi[\Omega]=\phi_0\frac{\Omega\tau}{1+(\Omega\tau)^2},
\end{equation}
although not frequency independent, is slowly varying around $\Omega \approx \tau^{-1}$, consistent with 
observations, and resolves the pathologies of a truly frequency-independent loss angle.
Since the $1/\Omega$ low-frequeny divergence of the displacement spectrum is mollified, most of the 
thermal energy of the oscillator is concerntrated around resonance; thus the approximation
$\chi[\Omega]^{-1}\approx m\left(2\Omega_0(-\Omega+\Omega_0)+i\Gamma_0\Omega_0\right)$
is viable (here $\Gamma_0=\Omega_0 \phi[\Omega_0]$). 

\begin{figure}[t!]
    \centering\includegraphics[width=\columnwidth]{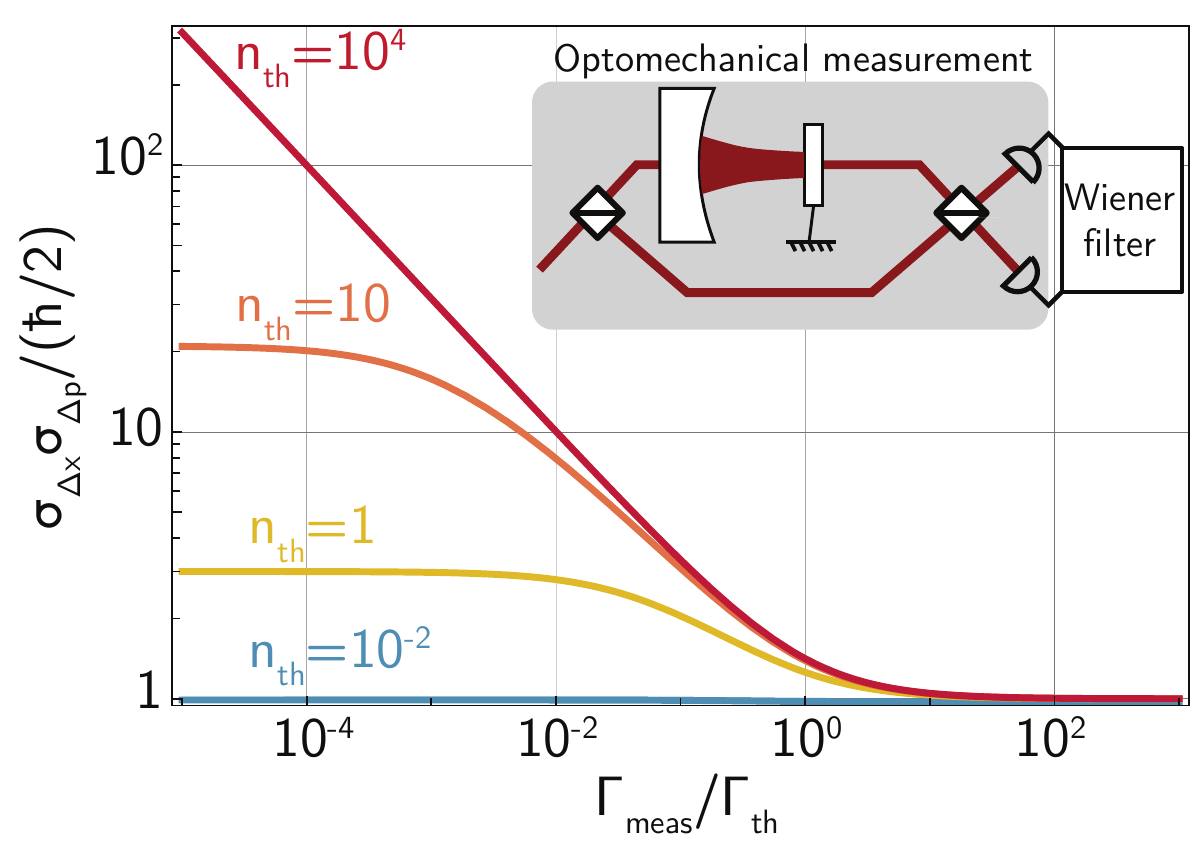}
    \caption{\label{fig:opto}
    Product of the uncertainty in the errors of the causal estimates of displacement and momentum of 
    a structuraly damped oscillator estimated from an interferometric measurement (inset).}
\end{figure}

We consider that the motion of the Zener-damped oscillator is measured using a cavity interferometer \cite{AspKip14}, 
as shown in the insert of \cref{fig:opto}. 
The motion of the oscillator changes the resonance frequency of the cavity, and thus the phase of light leaking out. 
Homodyne measurement of the light produces a photocurrent that is linearly proportional to the 
oscillator's displacement, together with detection noise; we denote by $\hat y$ these photocurrent fluctuations referred to apparent displacement.
Applying a causal Wiener filter to $\hat y$ gives an estimate of the physical displacement $\hat{x}$. The 
physical momentum $\hat{p}$ is then (using the close-to-resonance approximation), $\hat{p}[\Omega]=im\Omega_0\hat{x}[\Omega]$.
The estimation errors $\Delta\hat{x}$ and $\Delta\hat{p}$ are related similarly. 
The spectra of the errors in the estimate produced by causal Wiener filtering is \cite{Kailath1981,PhysRevA.80.043802}
\begin{equation}\label{eq:Sdxdx}
    S_{\Delta x_\ell\Delta x_{\ell'}}=S_{x_\ell x_{\ell'}} - 
    \left[\frac{S_{x_\ell y}}{S_y^-}\right]_+\left[\frac{S_{x_{\ell'} y}}{S_y^-}\right]_+^*,
\end{equation}
where $x_\ell \in \{\hat{x}$ or $\hat{p}\}$, and $S_{x_\ell x_{\ell'}}$ is the spectrum of the corresponding observables, 
$S_y^-$ is the anti-causal factor of the measured spectrum $S_{yy}$, such that $S_{yy}=S_y^+S_y^-$ with $S_y^+=\{S_y^-\}^*$, 
and $\left[.\right]_+$ takes the causal components of the expression in the bracket. 
The variance in the estimate $\Delta x_\ell$ is the integral of the spectrum $S_{\Delta x_\ell \Delta x_\ell}$.

For measurement using a cavity-interferometer, in the bad cavity regime 
(i.e. cavity linewidth $\kappa\gg\Omega_0$), with probe laser on-resonance with the cavity and 
homodyne detection of phase quadrature of the probe laser, direct computation using 
standard techniques \cite{AspKip14} shows that:
\begin{align}
    S_{xx}[\Omega] &= 
        \frac{2x_\mathrm{zpf}^2(\Gamma_\mathrm{th}+\Gamma_\mathrm{meas})}{(\Omega-\Omega_0)^2+(\Gamma_0/2)^2}
        = S_{xy}[\Omega]\label{eq:Sxx}\\
    \left[\frac{S_{xy}}{S_y^-}\right]_+ &=
    i\frac{2x_\mathrm{zpf}\sqrt{2\Gamma_\mathrm{meas}}(\Gamma_\mathrm{th}+\Gamma_\mathrm{meas})}
    {(\Gamma_0/2+\Gamma_\t{W})(\Omega-\Omega_0+i\Gamma_0/2)}
    \label{eq:+}
\end{align}
where $x_\mathrm{zpf}=\sqrt{\hbar/2m\Omega_0}$ is the zero-point fluctuation in displacement, $\Gamma_\mathrm{th}=\Gamma_0(n_\mathrm{th}+1/2)$ is the thermal decoherence rate, and
$\Gamma_\mathrm{meas}=4g^2/\kappa \equiv n_\t{meas} \Gamma_0$ is the measurement rate, where 
$n_\t{meas}=4g^2/(\kappa \Gamma_0)$ is the occupation due to quantum back-action of the measurement, 
and $g$ the multi-photon optomechanical coupling rate. 
$\Gamma_\t{W}$ is the characteristic estimation bandwidth of the Wiener filter, given by
$\Gamma_\t{W}^2 = 4\Gamma_\mathrm{th}\Gamma_\mathrm{meas}+4\Gamma_\mathrm{meas}^2+(\Gamma_0/2)^2$.
The spectrum of the estimation error [\cref{eq:Sdxdx}] is obtained by using
\cref{eq:Sxx,eq:+}:
\begin{equation}
\begin{split}
    S_{\Delta x\Delta x}[\Omega]=&
    \frac{
    8x_\mathrm{zpf}^2(\Gamma_\mathrm{meas}+\Gamma_\mathrm{th})}{\Gamma_0^2+4(\Omega-\Omega_0)^2}\\
    &\times \left[1-\frac{16 \Gamma_\t{meas}(\Gamma_\t{meas} + \Gamma_\t{th})}{(\Gamma_0 + \Gamma_\t{W})^2}\right].
\end{split}
\end{equation}
Integrating it gives the variance
\begin{equation}\label{eq:sigmaxBound}
    \sigma_{\Delta x}^2 = 
    x_\mathrm{zpf}^2\frac{4(\Gamma_\mathrm{meas}+\Gamma_\mathrm{th})}
    {\Gamma_0+2 \Gamma_\t{W}}.
\end{equation}
A similar computation for the momentum gives
\begin{equation}
    \sigma_{\Delta p}^2 = p_\t{zpf}^2 \frac{4(\Gamma_\mathrm{meas}+\Gamma_\mathrm{th})}
    {\Gamma_0+2 \Gamma_\t{W}},
\end{equation}
where $p_\t{zpf} = (\hbar/2)/x_\t{zpf}$ is the zero-point fluctuation in the momentum.
In both cases, straightforward analysis of the right-hand side shows that the variances 
are bounded as
\begin{equation}
\begin{split}
    (2n_\t{th}+1)x_\t{zpf}^2 &\geq \sigma_{\Delta x}^2 \geq x_\mathrm{zpf}^2 \\
    (2n_\t{th}+1)p_\t{zpf}^2 &\geq \sigma_{\Delta p}^2 \geq p_\mathrm{zpf}^2;
\end{split}
\end{equation}
here the lower bound is attained in the regime of strong measurement 
(i.e. $\Gamma_\t{meas} \gg \Gamma_\t{th}$), which is favourable for estimation,
while the upper bound is attained in the opposite regime. 
It is worth noting that even though quantum backaction contributes to the physical motion of the oscillator, 
as indicated by $\Gamma_\t{meas}$ in \cref{eq:Sxx}, 
the estimated motion can be free from it in the strong measurement regime for unity detection efficiency. 
Clearly the variances of the estimate errors satisfy
\begin{equation}
    (2n_\t{th}+1)^2 \frac{\hbar^2}{4} \geq \sigma_{\Delta x}^2 \sigma_{\Delta p}^2 \geq\frac{\hbar^2}{4},
\end{equation}
where the lower bound is the claim in \cref{eq:upxp}. 
\Cref{fig:opto} shows how the product $\sigma_{\Delta x} \sigma_{\Delta p}$ is bounded as 
the measurement rate $\Gamma_\t{meas}$ is increased.

\emph{Conclusion. }
We have proved --- with or without feedback, and without invoking any Markovian or Gaussian character of the system dynamics or measurement --- 
that the errors in causal estimation of quantum observables from a linear continuous measurement 
respects the Heisenberg uncertainty principle for the corresponding physical observables. 
We utilize linear response theory, widely applicable to current experimental measurement-based quantum control.
Furthermore, in the scenario with feedback control, we clarify that despite ``noise squashing'', 
the in-loop measurement record can provide as faithful an estimate of an observable as the 
out-of-loop record. As a matter of practice, this vastly simplifies experiments and extends the reach of 
quantum state estimation to non-Markovian scenarios \cite{groeblacher2015observation,meng2022measurement,whittle2021approaching}. 
Importantly, this insight eliminates the compromise in measurement efficiency
that is required in having two simultaneous measurements on the same system.\\

\noindent \emph{Acknowledgements. } We wish to thank C. Meng and W. Bowen for fruitful discussions. We acknowledge the generous support from the National Science Foundation (NSF). 
Especially, JC and BBL are supported through the NSF grant PHY-20122088, DG by the NSF grant PHY-0823459, and SD and YC by  NSF grants PHY-2011961, PHY-2011968, and PHY--1836809. In addition SD and YC acknowledge the support by the Simons Foundation (Award Number 568762).

\bibliography{main}

\end{document}